\documentstyle[aps,preprint]{revtex}
\begin{document}
\preprint{UVA-INPP-95/3}

\title{Accidental $\xi$-scaling as a Signature of
Nuclear Effects at $x>1$}

\author{Omar Benhar$^{\dagger}$ and Simonetta Liuti$^{* \dagger}$}

\address{$^{\dagger}$
INFN, Sezione Sanit\`a. Physics Laboratory, Istituto Superiore
di Sanit\`a.\\ Viale Regina Elena, 299. I-00161 Rome, Italy. \\
$^{*}$
Institute of Nuclear and Particle Physics, University of
Virginia.\\Charlottesville, Virginia 22901, USA.}

\maketitle

\begin{abstract}
We propose an interpretation of the
$\xi$-scaling behavior of nuclear structure functions
observed at Bjorken $x>1$ and $Q^2 \lesssim 4 \, {\rm (GeV/c)^2}$.
We show that at $\xi \gtrsim 1$, $\xi$-scaling might
arise accidentally because of the approximate cancellation of
two different $Q^2$-dependent effects, namely
Final State Interactions and
the effects implicit in the choice of the scaling variable $\xi$.
We provide a new convolution formula for the nuclear structure function
in terms of $\xi$ and make predictions
for the kinematical regions where Final State Interactions are expected to
be small and the suggested balancing of scaling violations is expected
to break down.
Our analysis is aimed at the final goal of clarifying the range of
applicability of local duality ideas in nuclei.
\end{abstract}
\narrowtext
For many years inclusive scattering of high energy leptons from nuclei has
been providing a
continuous flow of information both on nuclear dynamics
at short distances and on the internal structure of the nucleon.
With the experimental discovery of the European Muon Collaboration
(EMC) effect
it has become clear that a better understanding of the mechanisms that
modify quarks and gluons distributions
inside nuclei would help unraveling
unknown aspects of the dynamics of strong interactions.
In nuclei one can explore the kinematical regime beyond Bjorken $x=1$
($x=Q^2/2M_N \nu$, $Q^2$ being the four-momentum transfer
squared and $\nu$ the energy transfer)
where partons carry a higher momentum than in a single nucleon.
In dynamical models one expects deep inelastic nuclear structure
functions at $x \gtrsim 1$ not to vanish because of the
existence of unusual configurations
of the nucleus in which spectator particles
are directly involved in the scattering process.
A quite successful effective description of
such configurations is given for example
by nucleon-nucleon (NN) correlations
viewed as two close nucleons strongly recoiling against each other.
Unfortunately the cross sections in this kinematical domain fall very
steeply, making the extraction of nuclear structure functions
very challenging
(published data at $Q^2 > 50 \; {\rm (GeV/c)^2}$ and $x>1.15$
\cite{BCDMS,CDHSW}
only represent upper limits for the nuclear
structure functions).
Recently \cite{Fil,proposal}, it was suggested that
an indirect experimental access to the deep inelastic structure
functions could be obtained by
exploiting the connection between the low $Q^2$
regime and the asymptotic limit, known as Bloom and Gilman duality
\cite{BloGil}.
As a matter of fact,
the nuclear structure function per nucleon, $\nu W_2^A/A$,
extracted from the nuclear data on $^4He$, $^{12}C$ and
$^{56}Fe$ in the region
$1 \leq Q^2 \leq 4 \; {\rm (GeV/c)^2}$, was found
to scale in Nachtmann's variable $\xi$ for $\xi < 1$, with
some relatively small scaling violations at larger $\xi$.
The resulting $\xi$-scaling
curve was suggested to be {\it consistent} with the high $Q^2$
structure function, therefore supporting the applicability of
duality ideas to nuclei \cite{Fil}.

In this paper we propose an explanation for the $\xi$-scaling behavior
of nuclear structure functions at low $Q^2$ and high Bjorken $x$
by making a connection
with scaling in West's variable $y$ \cite{West}.
Our main aim is to try to clarify a
possible accidental nature of the scaling in $\xi$.
The removal of this ambiguity seems rather urgent
to us in view of the forthcoming experiments at the
Continuous Electron Beam Accelerator Facility (CEBAF) \cite{proposal}.

We show that at high $x$ ($\xi \gtrsim 1$),
where scattering occurs prevalently in the
Quasi Elastic (QE) channel, $\xi$-scaling is approached from below
as a result of the compensation of two opposite effects.
In fact it is a well known prediction of calculations based on
Impulse Approximation (IA),
that QE data should exhibit $y$-scaling
at high  $Q^2$
(for a recent review on $y$-scaling
see {\it e.g.} \cite{Donetal}).
At the $Q^2$ values of the present data
$y$-scaling is indeed observed in proximity of the QE peak
($y \approx 0$, $x \approx 1$) and
scaling violations
due to Final State Interaction (FSI) effects
set in as the energy transfer gets closer
to its threshold, corresponding to
large negative $y$ and $x>1$.
These scaling violations produce an overall enhancement of the
cross section at negative $y$ with respect to the IA one
({\it i.e.} $y$-scaling is approached from above).
By studying the nuclear structure functions in
terms of the variable $\xi$,
one shatters the pattern of $y$-scaling violations
because the relationship between the variables
$y$ and $\xi$ is $Q^2$ dependent.
This $Q^2$ dependence might conspire, as we shall see,
to counterbalance FSI effects giving rise to a better scaling behaviour.

At lower values of $x$, corresponding to $\xi < 0.6$,
the cross section receives a
larger contribution from the inelastic
channels and scaling in $\xi$ follows from the scaling of the nucleon
structure function, assuming that the bound-nucleon structure
function's $Q^2$-dependence is not sensibly
modified inside the nuclear medium.
At $ 0.6 \leq \xi \leq 0.8-0.9$, we expect
both inelastic and quasi elastic channels to contribute to the
structure functions. Their separation is obviously model dependent.
However, according to our calculation,
at {\it e.g.} $\xi \approx 0.8$
inelastic channels start to give a significant contribution only
at the highest values of $Q^2$ of present data ($Q^2 \approx 3 \,
{\rm (Gev/c)^2}$)
and one can isolate a kinematical region dominated by
the quasi-elastic peak.
In the last part of this paper we will try to clarify the
question of why should the nuclear structure functions in this region
fall along a scaling-limit curve as the available data seem to show.

We begin by defining the
the nuclear structure function,
$F_2^A(\nu,Q^2)=\nu W_2^A(\nu,Q^2)$, in IA
as a convolution of
the structure function for the bound nucleon, $\widetilde{F}_2^N$,
with the spectral function, $P^N(\mid {\bf k} \mid,E)$, $N=p,n$
(see also Ref.\cite{CDL92}):
\begin{eqnarray}
F_2^A(\nu,Q^2)& = &\int d{\bf k} \int dE \, Z \, P^p(\mid {\bf k} \mid,E)
\nonumber \\ & & \times
\widetilde{F}_2^{p}((kq),Q^2,k^2) {\cal F} + \nonumber
\\ & &
{\rm ( \, similar \, terms \, for \, neutrons \,) },
\end{eqnarray}
where $Z$ is the number of protons;
$k_\mu \equiv (k_o=M_A-(M_{A-1}^* + {\bf k} ^2)^{1/2}
,{\bf k}=-{\bf P}_{A-1})$
is the bound nucleon four-momentum,
${\bf P}_{A-1}$ is the momentum of the $A-1$ nucleus and
$M_{A,(A-1)}$ are the nuclear masses,
$M_{A-1}^* = M_{A-1} + E^*$, $E^*$ being the excitation energy
of the $A-1$ nucleus;
$E$ is the nucleon removal energy,
defined as $E= E_{min} + E^*$, with
$E_{min} \equiv M_{A-1}+ M_N-M_A$.
${\cal F}$
is a kinematical factor determined
by the choice of the off-shellness extrapolation
from the free nucleon structure functions (see e.g. \cite{CDL92,Sau}).
${\cal F}$ approaches unity at large $Q^2$.

We now make the following change
of  variables:
$d^3k dk_o  \equiv dk k dE d\xi^\prime \,  J_\xi^A$, with
$\xi^\prime \equiv 2x^\prime/(1+\sqrt{1+4k^2x^\prime/Q^2})$,
and $x^\prime=Q^2/2(kq)$. $\xi^\prime$ and $x^\prime$ play the role of
Nachtmann $\xi=2x/(1+\sqrt{1+4M_N^2 x /Q^2})$, and Bjorken
$x$ for an off-shell nucleon.
$J_\xi^A$ is the jacobian of the transformation, whose form we will
specify below.
We consider for illustration the case of infinite Nuclear Matter
(NM)
where kinematics is simpler because one does not account
for the recoil kinetic energy of the $A-1$ system
and calculations of realistic spectral functions and FSI effects
are available \cite{Benetal91}.
$F_2^{NM}(\xi,Q^2)$ reads:
\begin{eqnarray}
F_2^{NM}(\xi,Q^2) & = & 2 \pi Z \int_0^1
d \xi ^\prime   \, \,  F_2^{p}(\xi^\prime,Q^2)
\int_{E_{min}}^\infty dE
\nonumber \\ & & \times
\int_{k_{min}(Q^2,\xi,\xi ^\prime,E)}
    ^{k_{max}(Q^2,\xi,\xi ^\prime,E)} d\mid {\bf k} \mid
\, \mid {\bf k} \mid  \,
P^p(\mid {\bf k} \mid,E) \nonumber \\
& & \times {\cal F}  \, J^{NM}_\xi(Q^2,\xi^\prime,\mid {\bf k} \mid ,E)
\\ & & + \nonumber
{\rm ( \, similar \, terms \, for \, neutrons \,) }.
\end{eqnarray}
with  $J_\xi^{NM}= 1/(2q)(k^2+Q^2 \xi^\prime)$.
In practical calculations we identify the off-shell nucleon structure
function, $\widetilde{F}_2^N((kq),Q^2,k^2)$,
in Eq.(1) with the on-shell one,
$F_2^N(\xi,Q^2)$,
calculated at $\xi =\xi^\prime$. This correspond to disregarding
the dependency upon
the invariant $k^2$ which would imply strong nuclear medium modifications
of the bound nucleon.
The integration limits on $\mid {\bf k} \mid$ are:
\begin{mathletters}
\begin{eqnarray}
k_{min}  & = &  \left| M_N  \left( 1-\frac{\xi}{\xi^\prime} \right)
- E \right| \\
k_{max}  &  = & M_N \left( 1+Q^2\frac{\xi}{\xi^\prime} \right)
- E \approx \infty.
\end{eqnarray}
\end{mathletters}
Eqs.(2) and (3) describe both QE
and inelastic scattering,
depending on the form of  $F_2^N$.
In particular, for QE scattering
$F_2^N$
is a linear combination of the nucleon elastic form factors times
the delta function: $\delta(\xi-\xi_p)$, with
$\xi_p \equiv 2/(1+\sqrt{1+4M_N^2/Q^2})$.

Nuclear structure functions extend into the $x > 1$ region
according to the amount of high momentum and energy components
in $P^N(k,E)$.
A quantitative determination of the relative contributions of QE and
inelastic channels at $x>1$ is obviously model dependent
(see e.g. \cite{CDL92,Benetal91}).
However, presently available  calculations
\cite{CDL92,Benetal91,Vary2}
indicate that at $Q^2 \leq 4 \, {\rm (GeV/c)^2}$, {\it i.e.} in the
range of current data, QE scattering overrides completeley
inelastic scattering;
as $Q^2$ increases, QE scattering dies off along with the nucleon
form factors and the relative importance of inelastic channels
increases until one reaches a region
where neither process is clearly dominating;
in the very high $Q^2$ limit ($Q^2 \gtrsim 20 {\rm (GeV/c)^2}$ for
$x\approx 1.5$), deep inelastic scattering finally dominates
the cross section.

In Figure 1 we show as an example our results at fixed $\xi$ and in the
$Q^2$ range covered by the data of \cite{Fil}.
Figures (1a-1c) correspond to three different kinematical regions:
(a) the region where inelastic channels are almost completeley
suppressed and IA breaks down;
(b) an intermediate region where inelastic channels start to contribute
significantly only at the highest $Q^2$ values;
(c) the region beyond the QE peak where inelastic channels are
expected to give the major contribution to the cross section.
In what follows we will demonstrate that
$\xi$-scaling can occur accidentally in regions (a) and (b).

We expect data to exhibit a transition from $y$-scaling
to $\xi$-scaling, proceeding from region (a) to (c).
$y$-scaling for the
QE {\em reduced} cross sections ({\it i.e.}
the nuclear cross sections divided by the single nucleon cross sections,
$d^2\sigma_{A}/d^2\sigma_N \equiv F(y,Q^2)$)
was predicted in the high $Q^2$ limit
under the hypotheses that
only nucleon degrees of freedom are participating in
the scattering process and that IA is valid \cite{West}.
$y$
is the minimum longitudinal momentum carried by the struck nucleon
assuming that the spectator
$A-1$ system recoils with no excitation energy
\cite{Donetal}.
The definition of
$y$ depends on both the target and the recoiling system masses
and at a given kinematics its value changes depending on the
nucleus \cite{Liu93}.
In nuclear matter $y$ is defined as:
\begin{equation}
y_{NM}= -q+ \sqrt{(\nu-E_{min})^2+2M_N(\nu-E_{min})} .
\end{equation}
$y$-scaling was observed in experiments
along with scaling violations at low $Q^2$,
interpreted as an effect of FSI \cite{Donetal}.
$y$ is related to $\xi$ through:
\begin{eqnarray}
y & = & y_0(\xi) - \frac{M_N^2}{2q} + O(1/q^2)
\nonumber \\
& \equiv & y_0(\xi) - \frac{M_N^3 \xi}{Q^2} + O(1/Q^4),
\end{eqnarray}
with $y_0(\xi) \equiv M_N(1-\xi) - E_{min}$.

By calculating the structure functions at a fixed $\xi$,
one introduces a spurious $Q^2$ dependence
coming from the relationship between
$y$ and $\xi$.
We now consider the possibility that the $Q^2$ dependence of FSI
effects can counterbalance it.
We first write an expression for
the reduced cross section which includes the effect of  FSI,
$F_{FSI}(y,Q^2)$, in terms of the $y$-scaling quantity,
$F_{IA}(y)$:
\footnote{For consistency with previuos literature \cite{Donetal}
we consider here reduced cross sections instead of
nuclear structure functions. The IA calculation for $F_2^{NM}$
scales in $y$ {\em relatively} to the calculation including FSI, even if
$F_2^{NM}$ obviously
does not scale {\em per se}, or when compared to the
reduced cross section.}
\begin{equation}
F_{FSI}(y,Q^2) = F_{IA}(y+b_{FSI}(y,Q^2)).
\label{Fy}
\end{equation}
Here we define a shift in the variable $y$, $b_{FSI}(y,Q^2)$,
which is the projection onto the $y$ axis
of the variation in the cross section due to FSI,
$\Delta(y,Q^2)= F_{FSI}(y,Q^2)-F_{IA}(y)$.
We can define such a shift because the following properties are valid
at $x>1$:
{\it i)} both $F_{FSI}(y,Q^2)$ and $F_{IA}(y)$ are monotonously
increasing
functions of $y$, and, {\it ii)} $F_{FSI}(y,Q^2) > F_{IA}(y)$.
We obtained $b_{FSI}(y,Q^2)$ numerically by calculating $F_{FSI}$
using the approach
of Ref.\cite{Benetal91} (see also \cite{BenLiuFSI}).

We now replace $y$ on the right hand side of Eq.(\ref{Fy}) with
the expression in Eq.(5):
\begin{equation}
F_{FSI}(y,Q^2) = F_{IA}(y_0(\xi)+a_{\xi}(Q^2)+b_{FSI}(y,Q^2)),
\label{Fy2}
\end{equation}
One can clearly see that
$F_{FSI}(y,Q^2)$ becomes a function of $y_0(\xi)$ only
and therefore it  exhibits $\xi$-scaling,  to the extent to which
$a_\xi$ and $b_{FSI}$ compensate for each other.

In Figure 2 we show the quantities
$a_\xi$ and $b_{FSI}$,
at the fixed value of $y=-0.4$ GeV/c (corresponding to $\xi \gtrsim 0.9$).
The dashed line corresponds to IA.
$a_\xi$ and $b_{FSI}$ have opposite sign and therefore
they generate ``scaling violations''
that tend to compensate for each other.
Moreover, $\mid a_\xi \mid >b_{FSI}$ particularly
at low $Q^2$ and this is precisely
why in the available data, scaling seems to be approached
from below.
For comparison in Figure 2 we also show the quantity $a_x$, defined as:
\begin{eqnarray}
y & = & M_N(1-x) - E_{min} - a_x(Q^2),  \\
a_x(Q^2) & = & \frac{M_N^3 x (1-x^2)}{Q^2} + O(1/Q^4).
\end{eqnarray}
At $x>1$, $a_x$ being of the same sign of $b_{FSI}$,
contributes to enhance the scaling violations
due to FSI and this is at the origin of the
large $x$-scaling violations reported in \cite{Fil}.
We would also like to point out that, as also shown in Figure 2,
while FSI is expected to become negligible
at $Q^2 \approx 6 \; {\rm (GeV/c)^2}$,
the difference between $y$ and $\xi$ persists up
to much higher $Q^2$ values.
As a further proof of the validity of our
argument we predict $\xi$-scaling violations
to persist even at very high $Q^2$ (in Figure 2 we show values of $Q^2$
as large as  $10 \, {\rm (GeV/c)^2}$, in the range of CEBAF experiments
\cite{proposal}).
As shown in  Figure 2 such violations will approach the scaling
behavior from below and
they will be of the same magnitude as the ones observed at lower $Q^2$.
It is interesting to add that a similar effect
to the one that we just dicsussed was referred to in \cite{DGP}
as a possible explanation for the $\omega^\prime$ scaling observed in
the earlier data on nucleon
structure functions (Ref.\cite{BloGil} and references therein).
The authors of \cite{DGP} indeed suggested that logarithmic corrections
were compensating for the $Q^2$ dependent relationship between
$\omega=1/x$ and $\omega^\prime=\omega+M_N^2/Q^2$.

We now turn to the $\xi \approx 0.8-0.9$ region (Figure (1b)).
One is studying here the nuclear structure functions
in proximity of the QE peak; FSI effects
are small and the mechanism proposed to explain $\xi$-scaling
in region (a) cannot be applied.
The QE peak is positioned at:
$\xi_{peak} \approx \xi_p (1- \langle E \rangle /M_N )$,
corresponding to $k_{min}=0$ in Eq.(3a) ($\langle E \rangle$ is the
average value of the removal energy  and we disregard a small
$Q^2$-dependent correction due to FSI). We notice that
due to the $Q^2$ dependence of $\xi_p$,
$\xi_{peak}$ increases with increasing $Q^2$.
The height of the peak can be readily obtained from Eq.(2):
\begin{equation}
H_{peak}  \equiv  F_2^{NM}(\xi_{peak})
 \propto  \sigma_2(Q^2) ,
\label{peak}
\end{equation}
with
\begin{eqnarray}
\sigma_2(Q^2) & \approx &\frac{Z}{A} \,
\frac{2 M_N}{\left( 1+\frac{Q^2}{4 M_N^2} \right) }
\left( \left[ G_E^p(Q^2) \right]^2 + \frac{Q^2}{4 M_N^2}
\left[ G_M^p(Q^2) \right]^2 \right) +  \nonumber \\  & &
{\rm ( \, similar \, terms \, for \, neutrons \,) },
\end{eqnarray}
and $G_E^N$ and $G_M^N$ being the usual nucleon electric and magnetic
form factors. Eq.(\ref{peak}) is a consequence
of the fact that one is integrating over the whole range of momentum
and energy in Eq.(2) and the nucleon spectral function is normalized to one:
$\int dE \int dk \, k^2 P(k,E) = 1$.
$H_{peak}$ decreases with $\xi \equiv \xi_{peak}$ according to:
\begin{equation}
\sigma_2 \equiv  \sigma_2
\left( Q^2\equiv \frac{4 M_N^2}{\left[ \frac{2}{\xi} \left(
1 - \frac{\langle E \rangle}{M_N}  \right) -1 \right]^2 -1  }
\right)
\label{hpeak}
\end{equation}
In Figure 3 we
show for illustration our IA calculation of the QE peaks in $^{56}Fe$
and deuteron at different values of $Q^2$
($ 1 \, {\rm (GeV/c)^2} < Q^2 < 10  \, {\rm (GeV/c)^2}$)
versus $\xi$. We also show for comparison the curve for
$H_{peak}$, corresponding to Eq.(\ref{hpeak}).
$H_{peak}$ falls short with respect to the data
at high $Q^2$ where in fact inelastic channels start to set in.
However $\xi$-scaling was observed in a very small range of $Q^2$
(corresponding to the QE peaks farthest to the left in Figure 3)
where it seems to be mainly a consequence of the
rather large smearing at the top of the peaks for Fe.
In other words, from Figure 3 it is evident that the occurence or not of
$\xi$-scaling depends on the way the peak is smeared: one might
expect a scaling effect in
a complex nucleus such as iron, and not in deuteron for instance, where
the smearing is smaller.
The smearing in turn
reflects well known features of nuclear dynamics {\it i.e.}
the shape of the nucleon momentum distribution at low momentum
($k \leq 300 {\rm MeV}$).
We cannot envisage any fundamental reason behind this scaling behavior.
Our observation can be tested by performing a similar analysis
as the one presented in \cite{Fil}
using the data on deuteron \cite{Ddata} and $^3He$ \cite{3Hedata}.
We would like to notice, however,
a more intriguing feature in this kinematical region, namely
the falloff of QE peaks at different $Q^2$ s
relative to
the Deep Inelastic Scattering (DIS) limiting curve (the dots in Figure 3).

The falloff of the QE peak along the ``theoretical" DIS curve suggests
an interpretation analogous to  Bloom and Gilman's duality \cite{Fil}.
However, we believe that duality ideas
should be phrased in a different way in a nucleus.

We would like to state clearly that here one is observing the interplay
between two different``resonance to background" relationships, namely
the usual parton-hadron duality \cite{BloGil,DGP}
for the bound nucleon structure function and the occurence of
scattering into channels in which the final $A-1$ system
either recoils coherently or it undergoes breakup. These breakup channels
are identified with the nuclear background; coherent recoil generates
the low momentum and energy peaks corresponding to the ground
state for the $A-1$ nucleus, followed by its shell model excitations.

Now, nuclear dynamics contributes to
$\xi_{peak}$ and $H_{peak}$, Eq.(\ref{peak}),
through the average value of the removal energy, $\langle E \rangle$
and
the normalization of the nucleon spectral function, respectively.
These quantities in turn depend
mostly on the low energy and momentum components of $P(k,E)$.
Therefore, if one were to extract the DIS contribution from the elastic
cross section, according to the duality prescription \cite{BloGil,DGP}, one
would not get any information on the short distance nuclear dynamics which
is expected to strongly contribute in this region.
As a matter of fact, from Eqs.(2) and (3) one can see that
$\xi$ limits the phase space allotted for the contribution
of the nucleon spectral function to the DIS structure function.
As a result, at high enough $\xi$ ($\xi \gtrsim 0.8$) the
continuous background of $F^N_2$ is folded with the high $k$ part
of the nucleon spectral function, that is with the
{\em nuclear background} obtained
when breakup configurations for the $A-1$ system are included.
As $\xi$ increases, only the highest $k$ components contribute, which
occur with a decreasing
probability and this is what determines the structure
function's falloff.

To summarize, the behavior of the
low $Q^2$ structure function at $\xi \approx 0.8$
is determined by the elastic nucleon cross section and
by the low momentum components of the nuclear spectral function.
The high $Q^2$ structure function is determined by
the DIS nucleon structure function folded with the
high momentum components of the spectral function.
Because we are dealing with different
parts of the nucleon spectral function
these to quantities cannot be related in a straightforward
way using the usual duality ideas.
Our point of view is illustrated  also in Figure 3 where we
compare the QE peak falloff (dashed line) with the deep inelastic
limit curve (dotted line).

Finally we would like to comment briefly on the region of
low $\xi$ ($\xi <0.6$). Here the scaling in $\xi$ should reflect
the scaling in the nucleon structure function, modulo nuclear effects
that we expect to be of the same size of the EMC-effect.

Our conclusions are that the $\xi$-scaling found in the high $x$ data
seems to be most likely an accident.
When the data are plotted vs. $\xi$
the presence
of FSI effects in nuclei is hidden by the $Q^2$-dependent
relationship between
Nachtmann's $\xi$ and West's $y$.
In the particular region of $Q^2$ explored so far,
the pattern of $\xi$-scaling and $\xi$-scaling violations
does not seem to allow any space for any further theoretical
speculation.
Our interpretation can be tested readily with the
forthcoming experiments at CEBAF that will extend measurements to higher
$Q^2$. Here we predict that if the mechanism that we suggest
is correct,
$\xi$-scaling violations should persist with a comparable
magnitude as the one seen
at lower $Q^2$.
At $x \approx 1$ a new aspect of duality ideas is envisaged.
We emphasize that duality ideas in nuclei
should be considered within a more general framework
which includes together with the resonance to background
behavior implicit in nucleon structure,
the resonance and background behavior of nucleons
in nuclei, generated by the presence of short distance
NN configurations.
A more accurate discerning of the underlying dynamics of hadronic
configurations participating in
electron-nucleus
scattering processes in the multi-GeV region and at $x>1$
is a prerequisite
in order to explore consistently the exciting new aspects of QCD
in this region.

We thank Donal Day for useful discussions on the data and
the Institute of Nuclear and Particle Physics at the University
of Virginia for hospitality during the completion of this paper.

\begin{figure}
\caption{Comparison of the theoretical structure function per nucleon
for Fe with the experimental data of \protect\cite{Fil} plotted
vs. $Q^2$ at different values of $\xi$. Solid lines: full calculation,
 including both the quasielastic and the inelastic channels and the
 effect of FSI; dashes: contribution of the inelastic channels.}
\end{figure}

\begin{figure}
\caption{Shift in $y$, $a_\xi$ (short-dashed line) and
$b_{FSI}$ (full line) contributing to Eq.(\protect\ref{Fy2}),
plotted vs. $Q^2$ at fixed $y_{NM}=-0.4 /, {\rm GeV/c}$.
For comparison, the term $a_x$, Eq.(8), is also shown (dots).}
\end{figure}

\begin{figure}
\caption{Different contributions to the nuclear structure
function in deuteron (a),
and $^{56}Fe$, (b), vs. $\xi$.
The short-dashed curves in (a) and (b) represent the QE peaks
calculated in IA for values of $Q^2$
in the range
$ 1 \, {\rm (GeV/c)^2} < Q^2 < 8  \, {\rm (GeV/c)^2}$,
are shown ).
The full lines in (a) and (b) are the deep inelastic limit
of Eq.(2).
The dashed line in (b) represents $H_{peak}$ (Eq.(\protect\ref{peak})).
The data in (b) correspond to one of the kinematics of
\protect\cite{Fil} where scaling at
low $\xi$ was reported.}
\end{figure}
\end{document}